\documentclass[10pt]{article}
\usepackage{latexsym}
\usepackage[dvips]{graphicx,color}
\newcommand{\be}{\begin{equation}}
\newcommand{\ee}{\end{equation}}
\def\n{\noindent}
\catcode `\@=11
\catcode `\@=12
\begin{document}
\begin{center}
\large{\bf {Cosmological Models of Universe with Variable Deceleration Parameter in Lyra's Manifold}} \\
\vspace{10mm}
\normalsize{Anirudh Pradhan $^1$, J. P. Shahi$^2$ and Chandra Bhan Singh$^3$} \\
\normalsize{$^1$Department of Mathematics, Hindu Post-graduate College, 
Zamania-232 331, Ghazipur, India.} \\
\normalsize{E-mail : pradhan@iucaa.ernet.in, acpradhan@yahoo.com} \\
\vspace{5mm}
\normalsize{$^2,^3$Department of Mathematics, Harish-Chandra Post-graduate 
College, Varanasi, U. P., India.} \\
\end{center}
\vspace{10mm}
\begin{abstract} 
FRW models of the universe have been studied in the cosmological theory 
based on Lyra's manifold. A new class of exact solutions has been obtained 
by considering a time dependent displacement field for variable deceleration 
parameter from which three models of the universe are derived (i) exponential (ii) 
polynomial and (iii) sinusoidal form respectively. The behaviour of these 
models of the universe are also discussed. Finally some possibilities of 
further problems and their investigations have been pointed out.
\end{abstract}
\n Key words : {Cosmology, FRW universe, Lyra geometry } \\
\n PACS No. : 98.80.-k, 98.80.Jk
\section{Introduction}
\vspace*{-0.5pt}
Einstein proposed his general theory of relativity, in which gravitation is
described in terms of geometry, and it motivated the geometrization of other 
physical fields. One of the first attempts in this direction was made by Weyl 
\cite{ref1} who proposed a more general theory in which  gravitation and 
electromagnetism is also described geometrically. However, this theory was 
never considered seriously as it was based on the non-integrability of length 
transfer. Later Lyra \cite{ref2} suggested a modification of Riemannian geometry
by introducing a gauge function which removes the non-integrability condition 
of the length of a vector under parallel transport. Subsequently, Sen \cite{ref3} 
\& Sen and Dunn \cite{ref4} proposed a new scalar tensor theory of gravitation. 
They constructed an analog of the Einstein field equations based on Lyra's 
geometry which in normal gauge may be written as
\begin{equation}
\label{eq1}
R_{ij} - \frac{1}{2} g_{ij} R + \frac{3}{2}\phi_i \phi_j - \frac{3}{4}
 g_{ij}\phi_k \phi^k = - 8\pi G T_{ij},
\end{equation}
where $\phi_i$ is the displacement vector and other symbols have their 
usual meaning as in Riemannian geometry.
\par
Halford \cite{ref5} pointed out that the constant displacement vector field 
$\phi_{i}$ in Lyra's geometry plays the role of a cosmological constant 
in the normal general relativistic treatment. Halford \cite{ref6} showed that 
the scalar-tensor treatment based on Lyra's geometry predicts the same effects, 
within observational limits, as in Einstein's theory. Several authors \cite{ref7} 
have studied cosmological models based on Lyra's geometry  with a constant 
displacement field vector. However, this restriction of the displacement field 
to be a constant is a coincidence and there is no a {\it priori}reason for it. 
Singh et al. \cite{ref8} have studied Bianchi type I, III, Kantowski-Sachs 
and a new class of models with a time dependent displacement field and have made 
a comparative study of Robertson-Walker models with a constant deceleration 
parameter in Einstein's theory with a cosmological terms and in the cosmological 
theory based on Lyra's geometry. Recently Pradhan et al. \cite{ref9} and 
Rahaman et al. \cite{ref10} have studied cosmological models based on Lyra's 
geometry with constant and time time dependent displacement field in different 
context.

 The Einstein's field equations are a coupled system of highly nonlinear
differential equations and we seek physical solutions to the field equations
for their applications in cosmology and astrophysics. In order to solve the 
field equations we normally assume a form for the matter content or that 
space-time admits killing vector symmetries \cite{ref11}. Solutions to the field 
equations may also be generated by applying a law of variation for Hubble's 
parameter which was proposed by Berman \cite{ref12}. In simplest case the 
Hubble law yields a constant value for the deceleration parameter. It is 
worth observing that most of the well-known models of Einstein's theory and 
Brans-Dicke theory with curvature parameter $k=0$, including inflationary models, 
are models with constant deceleration parameter. In earlier
literature cosmological models with a constant deceleration
parameter have been studied by several authors \cite{ref9,ref13}.
But redshift magnitude test has had a chequered history. During the 1960s 
and the 1970s, it was used to draw very categorical conclusions. The 
deceleration parameter $q_{0}$ was then claimed to lie between $0$ and $1$ 
and thus it was claimed that the universe is decelerating. Today's situation,
we feel, is hardly different. Observations (Knop et al. \cite{ref14};
Riess et al. \cite{ref15}) of Type Ia Supernovae (SNe) allow to probe the
expansion history of the universe. The main conclusion of these
observations is that the expansion of the universe is accelerating. So we 
can consider the cosmological models with variable cosmological term and 
deceleration parameter. The readers are advised to see the papers by 
Vishwakarma and Narlikar \cite{ref16} and Virey et al. \cite{ref17} and 
references therein for a review on the determination of the deceleration 
parameter from Supernovae data.

Recently Pradhan et al. \cite{ref18} have studied the universe with 
time dependent deceleration parameter in presence of perfect fluid.
Motivated by the recent results on the BOOMERANG experiment on
Cosmic Microwave Background Radiation (Bernardis \cite{ref20}), we wish 
to study a spatially flat cosmological model. In this paper, we have 
investigated spatially non-flat and flat cosmological models with a 
time dependent displacement field within the framework of Lyra's geometry. 
We have obtained exact solutions of the field equations of Sen \cite{ref3} 
by taking the deceleration parameter to be variable. This paper is organized 
as follows. The metric and the field equations are presented in Section 2. 
In Section 3 we deal with a general solution. The Sections 4, 5, and 6 deal 
with the three different cases for the solutions in exponential, polynomial 
and sinusoidal forms respectively. Finally in Section 7 concluding remarks 
are given.
\section {Field Equations}
\noindent
The time-like displacement vector $\phi_i$ in the equation (\ref{eq1}) 
is given by
\begin{equation}
\label{eq2}
\phi_i = (0, 0, 0, \beta(t)).
\end{equation}
The energy-momentum tensor in the presence of a perfect fluid has the form
\begin{equation}
\label{eq3}
T_{ij} = (\rho + {p}) u_i u_j - {p} g_{ij}.
\end{equation}
together with co-moving coordinates $u^i u_i =1$, where $u_i = 
(0, 0, 0, 1)$.
The metric for FRW spacetime is
\begin{equation}
\label{eq4}
ds^2 = dt^{2} - R^{2}(t)\left[\frac{dr^{2}}{1 - kr^{2}} + r^{2}(d\theta^{2} 
+ \sin^{2}\theta d\phi^{2})\right] 
\end{equation}
where, $k = 1, -1, 0$. For this metric, the field equations (\ref{eq1}) with 
the equations (\ref{eq2}) and (\ref{eq3}) take the form 
\begin{equation}
\label{eq5}
3H^{2} + \frac{3k}{R^{2}} - \frac{3\beta^{2}}{4} = \chi \rho, 
\end{equation}
\begin{equation}
\label{eq6}
2\dot{H} + 3H^{2} + \frac{k}{R^{2}} + \frac{3\beta^{2}}{4} = - \chi p,
\end{equation}
where $\chi = 8\pi G$ and $H = \dot{R}/R$ is the Hubble's function. Equations 
(\ref{eq5}) and (\ref{eq6}) lead to the continuity equation
\begin{equation}
\label{eq7}
\chi \dot{\rho} + \frac{3}{2}\beta \dot{\beta} + 3\left[\chi(\rho + p) 
+ \frac{3}{2}\beta^{2}\right]H = 0.
\end{equation}
Assuming an equation of state
\begin{equation}
\label{eq8}
p = \gamma \rho, ~ ~ 0 \leq \gamma \leq 1
\end{equation}
Eliminating $\rho(t)$ from (\ref{eq5}) and (\ref{eq6}) we obtain
\begin{equation}
\label{eq9}
2\dot{H} + 3(1 + \gamma)H^{2} + (1 + 3\gamma)\frac{k}{R^{2}} + \frac{3}{4}
(1 - \gamma)\beta^{2} = 0.
\end{equation}
Here $\beta^{2}$ plays the role of a variable cosmological term $\Lambda(t)$. 
We have two independent equations in three unknown viz $R(t)$, $\rho(t)$ and 
$\beta$. Therefore we need one more relation among the variables in order to 
obtain a unique solution. Hence, we consider the deceleration parameter to be 
time dependent.  

\section{Solutions of the field equations}

We consider the deceleration parameter to be variable 
\begin{equation}
\label{eq10}
q = - \frac{R \ddot R}{\dot R^2} = - \left(\frac{\dot H + H^2}{H^2}\right) = 
b ~ (\rm variable).
\end{equation}
The equation (\ref{eq10}) may  be rewritten as 
\begin{equation}
\label{eq11}
\frac{\ddot R}{R} + b \frac{\dot R^2}{R^2} = 0.
\end{equation}
The general solution of Eq. (\ref{eq11}) is given by
\begin{equation}
\label{eq12} \int{e^{\int{\frac{b}{R}dR}}}dR = t + m,
\end{equation}
where $m$ is an integrating constant.\\

In order to solve the problem completely, we have to choose 
$\int{\frac{b}{R}}dR$ in such a manner that Eq. (\ref{eq12})
be integrable.

Without any loss of generality, we consider
\begin{equation}
\label{eq13} \int{\frac{b}{R}dR} = {\rm ln} ~ {L(R)},
\end{equation}
which does not effect the nature of generality of solution. Hence
from Eqs. (\ref{eq12}) and (\ref{eq13}), we obtain
\begin{equation}
\label{eq14} \int{L(R)dR} = t + m.
\end{equation}
Of course the choice of $L(R)$, in Eq. (\ref{eq14}), is quite
arbitrary but, since we are looking for physically viable models
of the universe consistent with observations, we consider the
following case:
\section{Solution in the Exponential Form}
Let us consider $ L(R) = \frac{1}{k_{1}R}$, where $k_{1}$ is arbitrary 
constant.

In this case on integration of  Eq. (\ref{eq14}) gives the exact solution 
\begin{equation}
\label{eq15} R(t) = k_{2}e^{k_{1}t}, 
\end{equation}
where $k_{2}$ is an arbitrary constant. Using Eqs. (\ref{eq8}) and (\ref{eq14}) 
in (\ref{eq9}) and (\ref{eq5}) or (\ref{eq6}), we obtain expressions for 
displacement field $\beta$, pressure $p$ and energy density $\rho$ as 
\begin{equation}
\label{eq16} \beta^{2} = - \frac{4(1 + \gamma)k_{1}^{2}}{(1 - \gamma)} 
- \frac{4(1 + 3\gamma)k}{3(1 - \gamma )k_{2}^{2}e^{2k_{1}t}}, 
\end{equation} 
\begin{equation}
\label{eq17} \chi p = \chi \gamma \rho = \frac{2\gamma}{(1 - \gamma)}\left[
3k_{1}^{2} + \frac{2k}{k_{2}^{2}e^{2k_{1}t}}\right].
\end{equation}
\begin{figure}[htbp]
\centering
\includegraphics[width=8cm,height=8cm,angle=0]{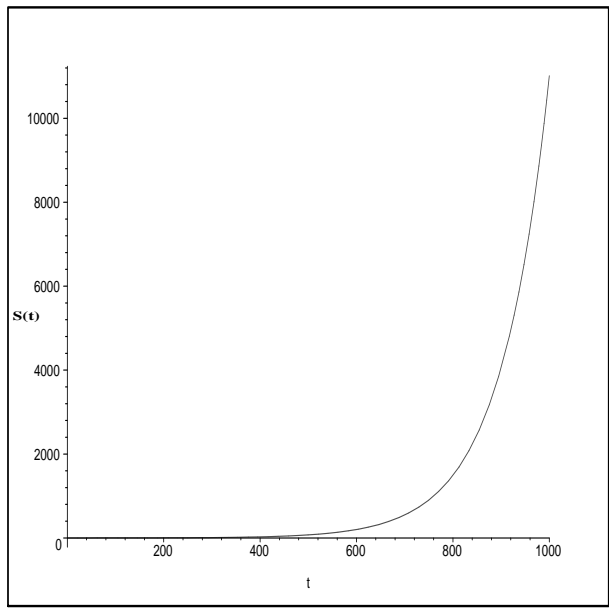}
\caption{The plot of scale factor $R(t)$ vs time with parameters 
$k_1 = 0.01$, $k_2 = 0.5$, and $\gamma = 0.5$}
\end{figure}
From Eq. (\ref{eq15}), since scale factor can not be negative, we find 
$R(t)$ is positive if $k_{2} > 0$. From Figure 1, it can be seen that 
in the early stages of the universe, i.e., near $t = 0$, the scale 
factor of the universe had been approximately constant and had increased 
very slowly. At specific time the universe had exploded suddenly and 
expanded to large scale. This is consistent with Big Bang scenario. \\

From Eq. (\ref{eq16}), it is observed that $\beta^{2}$ is a decreasing 
function of time. As mentioned earlier the constant vector displacement 
field $\phi_i $ in Lyra's geometry plays the role of cosmological constant 
$\Lambda$ in the normal general relativistic treatment and the scalar-tensor 
treatment based on Lyra's geometry predicts the same effects, within 
observational limits, as the Einstein's theory. Recent cosmological 
observations (Garnavich et al. \cite{ref20}, Perlmutter et al. \cite{ref21}, 
Riess et al. \cite{ref22}, Schmidt et al. \cite{ref23}) suggest the 
existence of a positive cosmological constant $\Lambda$ with the 
magnitude $\Lambda(G\hbar/c^{3}\approx 10^{-123}$. These observations on 
magnitude and red-shift of type Ia supernova suggest that our universe 
may be an accelerating one with induced cosmological density through the 
cosmological $\Lambda$-term. In our model, it is seen that $\beta$ plays 
the same role as cosmological constant and preserves the same character 
as $\Lambda$-term, in turn with recent observations. \\

From Eq. (\ref{eq17}), we observe that $p > 0$ and $\rho > 0$ for $k > 0$.
We also see that the energy density decreases to a small positive value and 
remains constant thereafter.
The expressions for $\beta^{2}$ and $\rho$ cannot be determined for the 
stiff matter $(p = \rho)$ models.
The expressions for $\beta^2$ and $\rho$ corresponding to
$\gamma = 0, 1/3$ are given in {\bf Table 1}. 

\vspace{0.5cm}

\begin{tabular}{|c|c|c|} \hline
$\gamma$        &   $\beta^2$                                     & $\rho$     \\ \hline
0               &   $-4[k_{1}^{2}+\frac{k}{3k_{2}^{2}e^{2k_{1}t}}]$   & $\frac{2}
{\chi}[3k_{1}^{2} + \frac{2k}{k_{2}^{2}e^{2k_{1}t}}]$    \\  \hline
$\frac{1}{3}$    &  $-4[2k_{1}^{2}+\frac{k}{k_{2}^{2}e^{2k_{1}t}}]$  & $\frac{3}
{\chi}[3k_{1}^{2} + \frac{2k}{k_{2}^{2}e^{2k_{1}t}}]$    \\  \hline    
\end{tabular} 
\vspace{0.5cm}

Table 1: Values of $\beta^2$ and $\rho$ for dust and radiation
exponential models. \\

We can obtain the values of $\beta^{2}$ and $\rho$ for {\bf flat FRW} model if 
we set $k = 0$ in Eqs. (\ref{eq16}) and (\ref{eq17}). 

\section{Solution in the Polynomial Form}
Let $ L(R) = \frac{1}{2k_{3}\sqrt{R + k_{4}}}$, where $k_{3}$ and $k_{4}$ 
are  constants. \\ 
In this case, on integrating, Eq. (\ref{eq14}) gives the exact solution
\begin{equation}
\label{eq18}
R(t) = \alpha_{1}t^{2} + \alpha_{2}t + \alpha_{3},
\end{equation} 
where $\alpha_{1}$, $\alpha_{2}$ and $\alpha_{3}$ are arbitrary constants. 
Using Eqs. (\ref{eq8}) and (\ref{eq18}) in (\ref{eq9}) and (\ref{eq5}) or 
(\ref{eq6}), we obtain the expressions for displacement field $\beta$, 
pressure $p$ and energy density $\rho$ as 
\begin{equation}
\label{eq19} \beta^{2} = \frac{16\alpha_{1}[\alpha_{3} + (2 + 3\gamma)
\alpha_{1}t^{2} + (2 + 3\gamma)\alpha_{2}t] + 4(1 + 3\gamma)(\alpha_{2}^{2} + k)}
{3(\gamma - 1)(\alpha_{1}t^{2} + \alpha_{2} t + \alpha_{3})^{2}},
\end{equation} 
\begin{equation}
\label{eq20} \chi p = \chi \gamma \rho = \frac{4\gamma[5\alpha_{1}^{2}t^{2} + 
5 \alpha_{1} \alpha_{2}t + \alpha_{2}^{2} + \alpha_{1} \alpha_{3} + k]}{(1 - \gamma)
(\alpha_{1}t^{2} + \alpha_{2} t + \alpha_{3})^{2}}.   
\end{equation}
\begin{figure}[htbp]
\centering
\includegraphics[width=8cm,height=8cm,angle=0]{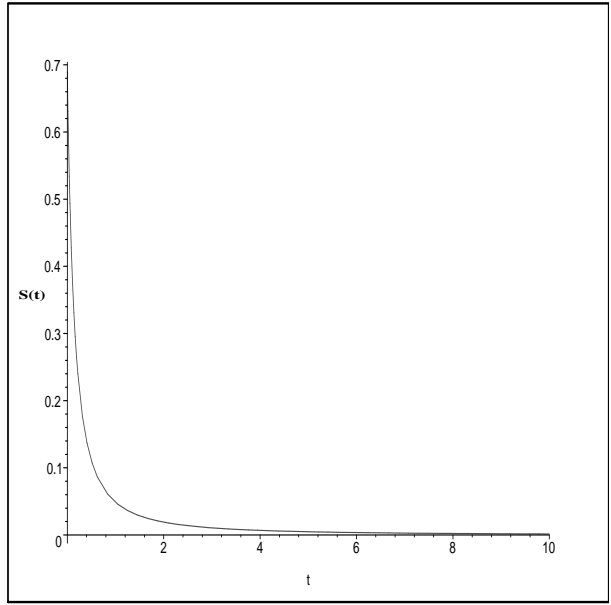}
\caption{The plot of scale factor $R(t)$ vs time with parameters 
$\alpha_1 = 1.00$, $\alpha_2 = 4.00$, $\alpha_3 = 1.00$ and $\gamma = 0.5$}
\end{figure}

From Eq. (\ref{eq18}), we note that $R(t) > 0$ for $0 \leq t < \infty$ 
if $\alpha_{1}$, $\alpha_{2}$ and $\alpha_{3}$ are positive constants.  
Figure $2$, shows that the scale factor is a decreasing function of 
time, implying that our universe is expanding. \\

Eq. (\ref{eq19}) shows that $\beta^{2} < 0$ for all times as 
$\gamma - 1 < 0$ and is a decreasing function of time, characteristically 
similar to $\Lambda$ in Einstein's theory of gravitation. In this model, 
$\beta$ plays the role as cosmological constant and it preserves the same 
character as $\Lambda$-term. This is consistent with recent observations 
(Garnavich et al. \cite{ref20}, Perlmutter et al. \cite{ref21}, Riess et al. 
\cite{ref22}, Schmidt et al. \cite{ref23}). A negative cosmological constant 
adds to the attractive gravity of matter; therefore, universe with a negative 
cosmological constant is invariably doomed to re-collapse. A positive 
cosmological constant resists the attractive gravity of matter due to 
its negative pressure. For most of the time, the positive cosmological constant 
eventually dominates over the attraction of matter and drives the universe to 
expand exponentially. \\

The expressions for $\beta^{2}$ and $\rho$ cannot be determined for the stiff 
matter $(p = \rho)$ models. For dust model $(\gamma = 0)$, $\beta^{2}$ and $\rho(t)$ 
are given by 
\begin{equation}
\label{eq21} \beta^{2} = -\frac{16\alpha_{1}[2\alpha_{1}t^{2} + 2\alpha_{2}t + \alpha_{3}] 
+ 4\alpha_{2}^{2} + 4k}{(\alpha_{1}t^{2} + \alpha_{2} t + \alpha_{3})^{2}},  
\end{equation}
\begin{equation}
\label{eq22} \chi \rho = \frac{4[5\alpha_{1}^{2} t^{2} + 5\alpha_{1}\alpha_{2}t + 
\alpha_{2}^{2} + \alpha_{1}\alpha_{3} + k]}{(\alpha_{1}t^{2} + \alpha_{2} t + \alpha_{3})^{2}}
\end{equation}
For radiative model $(\gamma = \frac{1}{3})$, $\beta^{2}$ and $\rho(t)$ 
are given by 
\begin{equation}
\label{eq23} \beta^{2} = -\frac{8\alpha_{1}[2\alpha_{1}t^{2} + 2\alpha_{2}t + \alpha_{3}] 
+ 8\alpha_{2}^{2} + 8k}{(\alpha_{1}t^{2} + \alpha_{2} t + \alpha_{3})^{2}},
\end{equation}
\begin{equation}
\label{eq24} \chi \rho = \frac{6[5\alpha_{1}^{2} t^{2} + 5\alpha_{1}\alpha_{2}t + 
\alpha_{2}^{2} + \alpha_{1}\alpha_{3} + k]}{(\alpha_{1}t^{2} + \alpha_{2} t + 
\alpha_{3})^{2}}
\end{equation}
If we set $k = 0$, in above equations (\ref{eq19}) - (\ref{eq24}), we get solutions 
for {\bf flat FRW} universe.

\section{Solution in the Sinusoidal Form}
Let $ L(R) = \frac{1}{\beta\sqrt{1 - R^{2}}}$, where $\beta$ is constant. \\
In this case, on integrating, Eq. (\ref{eq14}) gives the exact solution
\begin{equation}
\label{eq25}
R = M\sin(\beta t) + N\cos(\beta t) + \beta_{1},
\end{equation} 
where $M$, $N$ and $\beta_{1}$ are constants.
Using Eqs. (\ref{eq8}) and (\ref{eq25}) in (\ref{eq9}) and (\ref{eq5}) or 
(\ref{eq6}) , we obtain the expressions for displacement field $\beta$, 
pressure $p$ and energy density $\rho$ as 
\begin{equation}
\label{eq26} \beta^{2} = \frac{4[2\beta^{2}(M^{2} + N^{2}) + 2\beta^{2} \beta_{1}
(P - \beta_{1}) - 3(1 + \gamma)Q - (1 + 3\gamma)k]}{3(1 - \gamma)P^{2}},
\end{equation} 
\begin{equation}
\label{eq27} \chi p = \chi \gamma \rho = \frac{2\gamma[3Q + 2k - \beta^{2}(M^{2} + N^{2}) 
- \beta^{2}\beta_{1}(P - \beta_{1})]}{(1 - \gamma)P^{2}},   
\end{equation} 
where 
$$P = M\sin{\beta t} + N \cos{\beta t} + \beta_{1},$$
$$Q = (M\cos{\beta t} - N\sin{\beta t})^{2}.$$
\begin{figure}[htbp]
\centering
\includegraphics[width=8cm,height=8cm,angle=0]{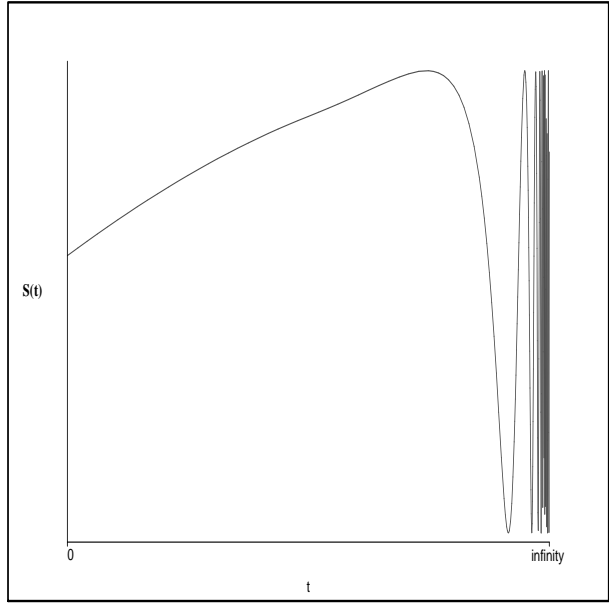}
\caption{The plot of scale factor $R(t)$ vs time with parameters 
$M = 2.00$, $N = 1.00$, $\beta = 10.00$, $\beta_1 = 0.2$, and $\gamma = 0.5$}
\end{figure}
From the Figure 3, we note that at early stage of the universe, the scale 
of the universe increases gently and then decreases sharply, and after wards it 
will oscillate for ever. We must mention here that the oscillation takes 
place in positive quadrant. This has physical meaning. \\

The expressions for $\beta^{2}$ and $\rho$ cannot be determined for the stiff 
matter $(p = \rho)$ models. For dust model $(\gamma = 0)$, $\beta^{2}$ and $\rho(t)$ 
are given by 
\begin{equation}
\label{eq28} \beta^{2} = \frac{4[2\beta^{2}(M^{2} + N^{2}) + 2\beta^{2} \beta_{1}
(P - \beta_{1}) - 3Q - k]}{3P^{2}}, 
\end{equation}
\begin{equation}
\label{eq29} \chi \rho = \frac{2[3Q + 2k - \beta^{2}(M^{2} + N^{2}) 
- \beta^{2}\beta_{1}(P - \beta_{1})]}{P^{2}}, 
\end{equation}
For radiative model $(\gamma = \frac{1}{3})$, $\beta^{2}$ and $\rho(t)$ 
are given by 
\begin{equation}
\label{eq30} \beta^{2} = \frac{4[\beta^{2}(M^{2} + N^{2}) + \beta^{2} \beta_{1}
(P - \beta_{1}) - 2Q - k]}{P^{2}}, 
\end{equation}
\begin{equation}
\label{eq31} \chi \rho = \frac{3[3Q + 2k - \beta^{2}(M^{2} + N^{2}) 
- \beta^{2}\beta_{1}(P - \beta_{1})]}{P^{2}},  
\end{equation}
For {\bf flat FRW} universe, we put $k = 0$ in above results. Since, in these cases, 
we have many alternatives for choosing values of $M$, $N$, $\beta$, $\beta_{1}$, 
it is just enough to look for suitable values of these parameters, such that 
the physical initial and boundary conditions are satisfied.
\section{Conclusions}
In this paper we have obtained exact solutions of Sen's equations in Lyra 
geometry for time dependent deceleration parameter in FRW spacetime. The 
nature of the displacement field $\beta(t)$ and the energy density $\rho(t)$ 
have been examined for three cases (i) exponential form (ii) polynomial 
form and (iii) sinusoidal form. The solutions obtained in Sections $4$, $5$ 
and $6$ are to our knowledge quit new. Here it is found that the displacement 
field plays the role of a variable cosmological term $\Lambda$. \\

Recently there is an upsurge of interest in scalar fields in general 
relativity and alternative theories of gravitation in the context of
inflationary cosmology (La and Steinhardt \cite{ref24} ; Ellis \cite{ref25}; 
Barrow \cite{ref26}). Therefore the study of cosmological models in Lyra 
geometry may be relevant for inflationary models. Further the space dependence 
of the displacement field $\beta$ is important for inhomogeneous models for 
the early stages of the evolution of the universe. Besides, the implication 
of Lyra's geometry for astrophysical interesting bodies is still an open 
questions. The problem of equation of motion and gravitational radiation need 
investigation. Finally in spite of very good possibility of Lyra's geometry 
to provide a theoretical foundation for relativistic gravitation, astrophysics 
and cosmology, the experimental point is yet to be undertaken. But still 
the theory needs a fair trial. 
\section*{Acknowledgements}
One of the authors (A. Pradhan) would like to thank the Inter-University 
Centre for Astronomy and Astrophysics, Pune, India for providing facility 
where part of the paper was carried out. The authors also thank to the referees 
for pointing out some typos.  

\end{document}